\documentclass[prd,onecolumn,notitlepage,
nofootinbib,aps,tightenlines,
preprintnumbers,amsmath,amssymb,amsfonts,showpacs,superscriptaddress]{revtex4-1}

\usepackage[hyperindex,breaklinks]{hyperref}
\usepackage{graphicx,epstopdf}
\usepackage[utf8]{inputenc}
\usepackage{epsf, array, color}
\usepackage{graphicx}
\usepackage{subfigure}
\usepackage{bbold}
\usepackage{slashed}
\usepackage{amsthm}
\usepackage{amsmath}
\renewcommand{\eqref}[1]{Eq.~(\ref{#1})}

\usepackage{natbib,bm}
\usepackage{hyperref}
\usepackage{breakurl}
\usepackage{multirow}
\usepackage{bbold}
\usepackage{listings}

\DeclareMathAlphabet{\mathcalligra}{T1}{calligra}{m}{n}

\DeclareMathAlphabet{\mathcalligra}{T1}{calligra}{m}{n}
\newcommand{\beq}{\begin{equation}}
\newcommand{\eeq}{\end{equation}}
\renewcommand{\d}{\text{d}}
\newcommand{\f}{\frac}
\newcommand{\be}{\begin{eqnarray}}
\newcommand{\ee}{\end{eqnarray}}
\newcommand{\bea}{\begin{eqnarray}}
\newcommand{\eea}{\end{eqnarray}}

\renewcommand{\d}{\text{d}}

\begin{document}
\title{Muon Beam Experiments to Probe the Dark Sector}

\author{Chien-Yi Chen}
\thanks{cchen@perimeterinstitute.ca}

\author{Maxim Pospelov}
\thanks{mpospelov@perimeterinstitute.ca}
\affiliation{Department of Physics and Astronomy, University of Victoria, Victoria, BC V8P 5C2, Canada}
\affiliation{Perimeter Institute for Theoretical Physics, Waterloo, ON N2J 2W9, Canada}

\author{Yi-Ming Zhong}
\thanks{ymzhong@bu.edu}
\affiliation{Physics Department, Boston University, Boston, 02215, USA}

\begin{abstract}
 A persistence of several anomalies in muon physics, such as the muon anomalous magnetic moment and the muonic hydrogen Lamb shift, hints at
new light particles beyond the Standard Model. We address a subset of these models that have a new light scalar state with sizable couplings to muons and suppressed couplings to electrons. 
A novel way to search for such particles would be through muon beam-dump experiments by (1) missing momentum searches; (2) searches for decays with displaced vertices. 
The muon beams available at CERN and Fermilab present
 attractive opportunities for exploring the new scalar with a mass below the di-muon threshold, and potentially 
covering a range of relevant candidate models. For the models considered in this paper, both types of signals, 
muon missing momentum and anomalous energy deposition at a distance, can probe a substantial fraction of the unexplored parameter space of the new light scalar, including a  region that can explain the muon anomalous magnetic moment discrepancy.   
\end{abstract}	
\maketitle

\section{Introduction}

 New Physics (NP) at low-mass, treated in all generality, has become an actively pursued topic of the intensity frontier physics \cite{Hewett:2012ns,Essig:2013lka,Alexander:2016aln} given the abundant evidence for NP in the neutrino and dark matter sectors, coupled with the lack of NP signal at the Large Hadron Collider (LHC).
Motivations for searches of low-mass, weakly-coupled particles can come from 
 top-down theoretical arguments (see {\em e.g.} \cite{Jaeckel:2010ni}). But a bigger role is played by 
the existing anomalous observations in particle experiments, 
astrophysics, and cosmology, which might find their explanations in models with  NP at low-mass
(see {\em e.g.} \cite{Boehm:2003hm,Pospelov:2007mp,ArkaniHamed:2008qn}).
The current $\sim 3.5\sigma$ discrepancy between the predicted and observed value of the muon anomalous magnetic moment
\cite{Bennett:2006fi}, $a_\mu$,  will 
be scrutinized in the upcoming experiments at Fermilab and JPARC \cite{Grange:2015fou,Mibe:2010zz}. 
It is not clear that the current tension is a result of experimental errors or  theoretical errors or a combination of the two. With new measurements of muon $g-2$ and improved Standard Model (SM) calculations 
based on lattice QCD \cite{Benayoun:2014tra, Jin:2016rmu}, one hopes to clarify the origin of the existing discrepancy. 
Lamb shifts of muonic atoms, such as muonic hydrogen and deuterium \cite{Pohl:2010zza,Antognini:1900ns,Pohl1:2016xoo}, 
present another formidable puzzle. When interpreted in terms 
of the charge radius of the proton, $r_p$, these measurements disagree with the electron scattering and hydrogen spectroscopy extracted values of $r_p$ by $\sim 7 \sigma$ \cite{Mohr:2015ccw}.

In this paper, we are interested in the scenarios where the deficit of 
theoretical predictions for $a_\mu$ is compensated by a contribution
from NP.  Although the overall size of the $a_\mu$ discrepancy, 
$a_\mu^{\rm obs} -  a_\mu^{\rm th} \approx +3\times 10^{-9} $,  is on the order of the 
corresponding contributions from the weak sector of the SM, the NP states correcting the anomalous magnetic moment do not have to reside at the 
weak scale. Indeed it is well known that the existing theoretical deficit can be compensated by loop contributions from new light particles \cite{Gninenko:2001hx,Fayet:2007ua,Pospelov:2008zw}. One such candidate model, the dark photon, has been searched for in a variety of experiments, with recent results ruling out the most minimal version
as a possible explanation of the $a_\mu$ discrepancy. Some other candidate models still survive the existing constraints, including 
the $L_\mu-L_\tau$ gauged model and its variations \cite{Altmannshofer:2014pba,Altmannshofer:2016brv}.

Here we would like to examine the models with a new light scalar, $S$, tuned to explain the $a_\mu$ discrepancy \cite{Chen:2015vqy,Batell:2016ove}. 
We will employ a simplified framework, with a relevant Lagrangian given by
\be
{\cal L}\supset{1\over2}(\partial_\mu S)^2 -{1\over2}m_S^2 S^2 - \sum_{\ell =e, \mu, \tau} g_\ell S \bar{\ell}\ell,
\label{simple}
\ee
where  $g_\ell$ is the coupling between $S$ and leptons. Notice that \eqref{simple} is an effective Lagrangian that does not respect the full gauge symmetry of the SM. 
Its $SU(2)\times U(1)$ generalization  is given by the following dimension-five effective operator,
\be
{\cal O}_5 = \frac{1}{\Lambda} (\bar L E) H S, 
\label{O}
\ee
where $H$ is the SM Higgs doublet, and $L$, $E$ are the lepton doublets and singlets respectively. The effective operator, $\mathcal O_5$,
can be embedded into a full model in a variety of ways. 
Refs. \cite{Chen:2015vqy,Batell:2016ove} discuss the phenomenology of such a model and choose different types 
of UV completion of $\mathcal O_5$ with vector-like fermions or multiple Higgs states respectively. The latter UV completion 
generates strong constraints for the mass range $m_S > 2 m_\mu$ due to recent searches of unexpected peaks in the 
di-muon mass spectrum for $B\to K(\mu^+\mu^-)$ decays at the LHCb \cite{Aaij:2015tna}. The mass range of $m_S<2 m_\mu$ remains 
largely unexplored. In this mass range, the new light particles can be relatively long-lived, and thus amenable to beam-dump experiments and fixed-target searches. 

\begin{figure}[t]
\centering
\includegraphics[width=0.5\columnwidth]{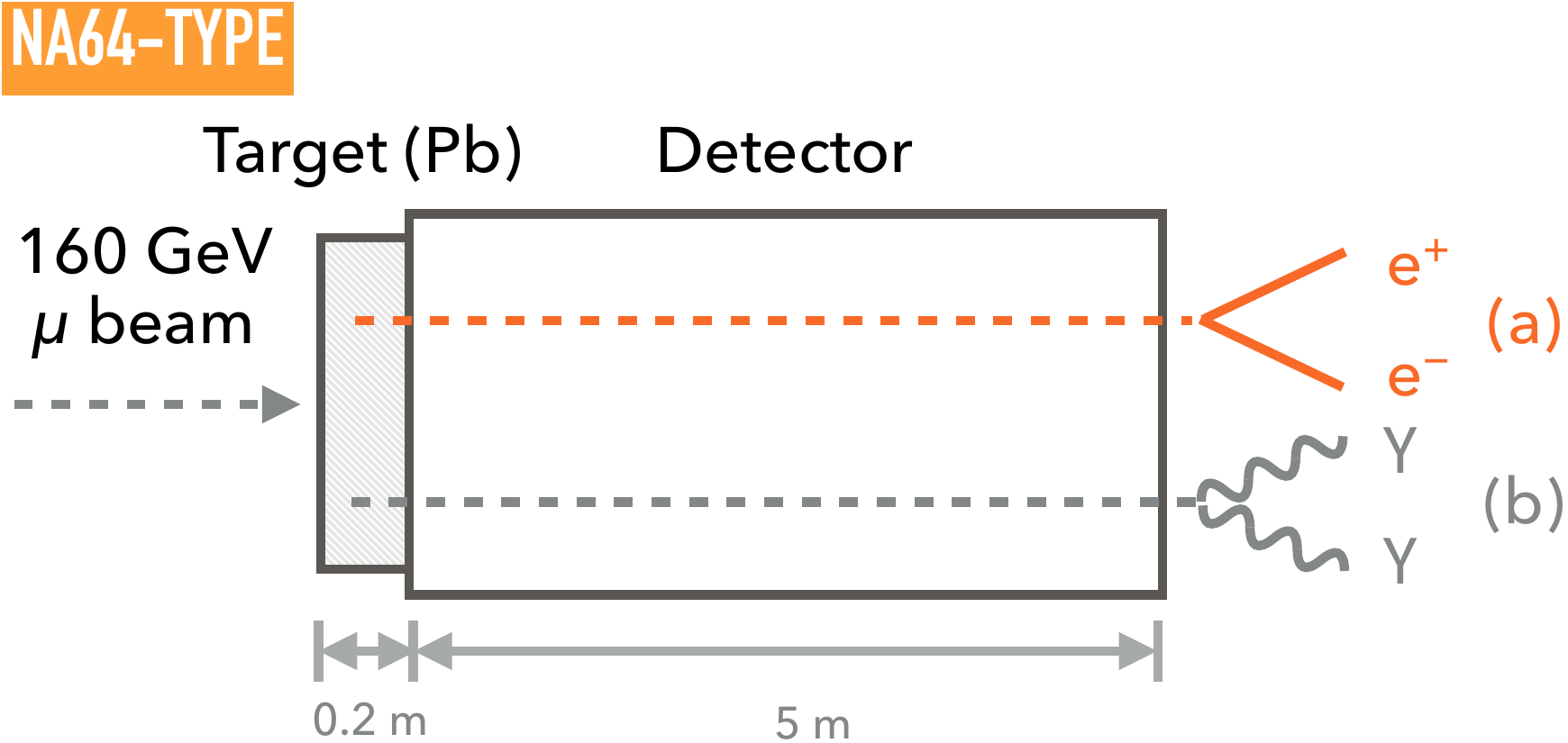}\hspace*{1em}~
\includegraphics[width=0.379\columnwidth]{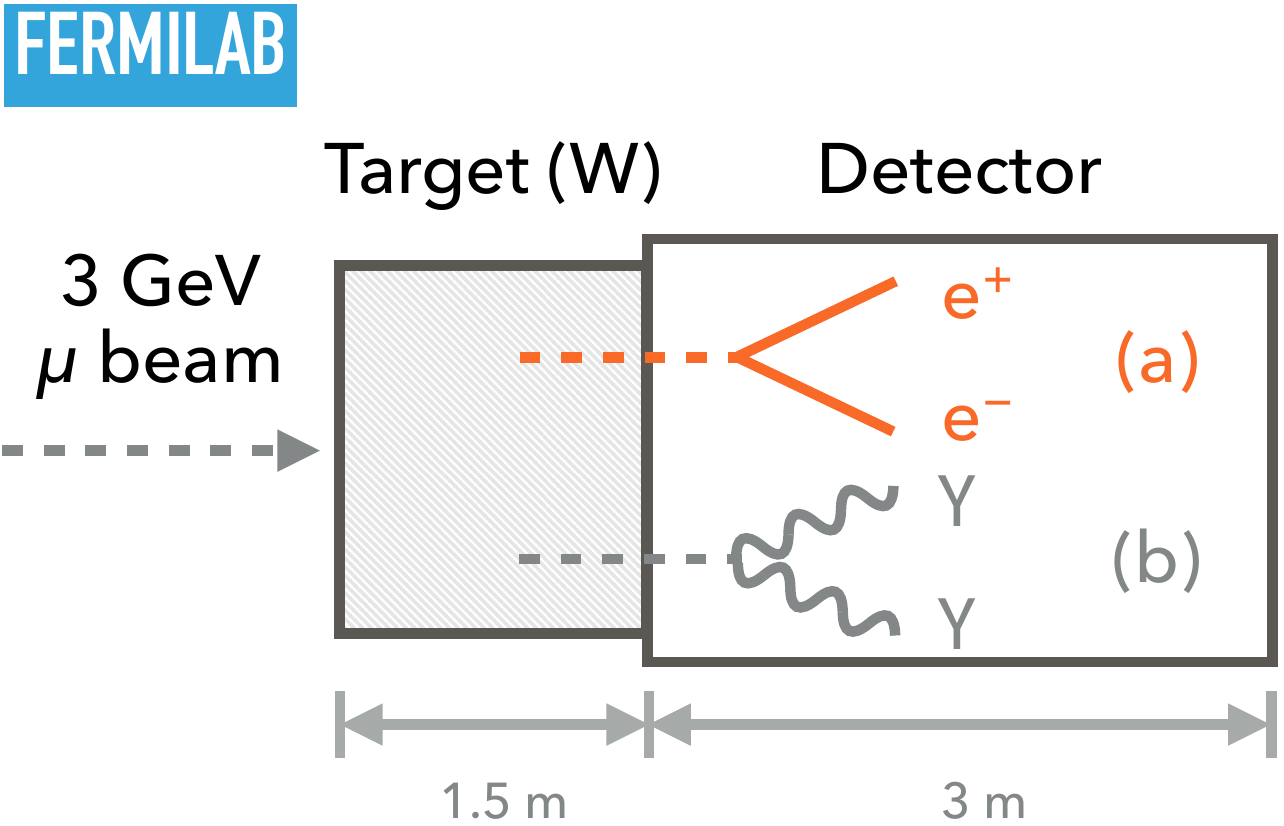}\hspace*{2em}
\caption{\label{fig:setup}Setups for muon beam-dump experiments at NA64 (left) and  Fermilab (right).  For the NA64-type experiment, the muon beam energy is $\sim$160 GeV and the target material is lead. We focus on the missing energy searches with $S$ decays into $e^+ e^-$ (a) and $\gamma\gamma$ (b). For Fermilab experiment, the muon beam energy is $\sim$3 GeV and the target material is tungsten. We focus on the decays with displaced vertices of $S$. The lengths of the targets and the detectors are shown in the plot.}
\end{figure}

\begin{figure}[t]
\centering
\includegraphics[width=0.35\columnwidth]{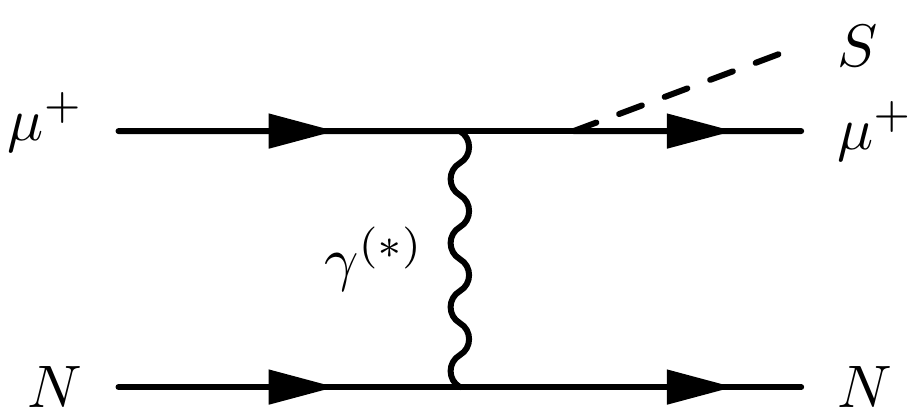}\hspace*{1em}~
\includegraphics[width=0.35\columnwidth]{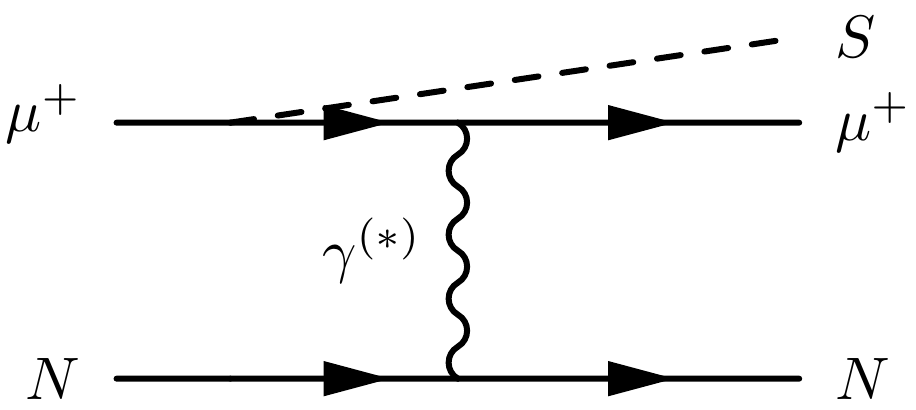}\hspace*{2em}
\caption{\label{fig:brem} Feynman diagrams illustrating the bremsstrahlung production of the new light states, $S$, where an incoming muon, $\mu^+$, interacts with a target nucleus, $N$, by exchanging a photon, $\gamma$, and radiates the exotic scalar, $S$.}
\end{figure}

In this paper, we investigate the potential of experiments where the light scalar, $S$, is sourced by the collision of 
muons with nuclei. Subsequent displaced decays of $S$ present an opportunity for both the 
missing momentum and the anomalous energy deposition searches. We take the existing CERN and Fermilab muon\footnote{``Muon" or ``$\mu$" refers to $\mu^+$ for the muon beam-dump experiments.} sources as an example and illustrated our main idea in Fig.~\ref{fig:setup}. For a NA64-type setup at CERN \cite{Andreas:2013lya,Gninenko:2014pea},  the \emph{dark} emission of $S$ states with $S$ decaying outside of the detector would cause the anomalous loss 
of muon energy, which can be detected in the muon scattering experiment.
The  muon beam with beam energy around a few GeV  at Fermilab would also provide a great opportunity. Here muons are stopped in the dense material, and subsequent anomalous energy deposition is searched 
directly behind it. In what follows, we  demonstrate that
both approaches  allow probing unexplored parts of the parameter space of the simplified model 
potentially responsible for the $a_\mu$ discrepancy. We use the bremsstrahlung, $\mu^+ + N \to \mu^+ + N + S$, as the main production mechanism as illustrated in Fig.~\ref{fig:brem}, where an incident muon, $\mu^+$, interacts with a target nucleus, $N$, by exchanging a photon, $\gamma$, and radiates the exotic scalar, $S$. The two muon beam-dump experiments considered in this paper can be easily implemented with only modest modifications/additions to the existing experimental infrastructure. Looking into more distant future, the proton beam-dump facilities, such as SHiP \cite{Alekhin:2015byh}, would also provide strong sensitivities to muon-coupled light states.

The paper is organized as follows:  We first show two specifications of the simplified model in Sec.~\ref{modelswithnewlightscalars} and analyze the two proposed muon beam-dump experiments in Sec.~\ref{signalsandbackgrounds}. We  show the resulting expected sensitivities in Sec.~\ref{results}, conclude, and discuss other related experiments in Sec.~\ref{conclusions}.

\section{Models with new light scalars}
\label{modelswithnewlightscalars}
For the simplified model introduced in the previous section, \eqref{simple}, the couplings $g_{\ell=e, \mu, \tau}$ are free parameters.  The muon anomalous magnetic moment, 
 $a_\mu$, receives corrections due to the one-loop contribution of $S$, 
\be
\Delta a_\mu = \frac{g_\mu^2}{8\pi^2}\int_0^1\d z \frac{(1-z)^2(1+z)}{(1-z)^2 + z (m_S/m_\mu)^2} .
\label{g-2}
\ee
Requiring this correction to reduce current tension between measurement and the SM calculation of 
$a_\mu$, one arrives at the preferred values of $\{g_\mu,m_S\}$ parameters. 
For example, given $g_\mu = 5\times 10^{-4} $ and $m_S = 100$~MeV, 
$\Delta a_\mu$ is around $1.6\times 10^{-9}$. This would bring the theoretical and experimental values for the muon anomalous magnetic moment within $2\sigma$.  Other couplings,  $g_e$ and $g_\tau$, still remains free.  Here we will consider 
two models with further  specifications on the couplings of $g_e$ and $g_\tau$:
\begin{itemize}
\item Model A:  Mass proportionality, $g_\ell \propto m_\ell$. In particular, it implies that the couplings between the scalar $S$ and  electrons are 
$\sim 200$ times smaller than those with muons. Despite this, the dominant decay channel for $S$ below the di-muon threshold  is $S\to e^+e^-$. 
\item Model B: Coupling exclusively to muons, $g_\mu \neq 0$ and $g_e=g_\tau = 0$. This is the most collider and electron/proton beam-dumps unfriendly
case, that still can be relevant for the muon $g-2$. Due to the longer lifetime of $S$, 
the  missing energy search at NA64-type experiments would be particularly useful to constrain the parameter space of 
this model. 
\end{itemize} 

Model A can be explicitly constructed using the leptonic Higgs doublet model completion of Ref. \cite{Batell:2016ove}. 
In that model, the lepton flavor conservation and $g_\ell\propto m_\ell$ proportionality are guaranteed, as only one Yukawa matrix 
determines the lepton masses and their couplings to $S$. Model B is in some sense more artificial but phenomenologically minimal. Only one coupling 
is introduced that is necessary to correct $a_\mu$ in this model. 

In both models,  $S$ dominantly decays to $\mu^+\mu^-$ for $2m_\mu < m_S < 2m_\tau$. For the mass range that we concentrate on, $2m_e< m_S<2m_\mu$, the total decay width, $\Gamma_S$, is a sum of the decay widths of  $S\to e^+e^-$ and $S\to \gamma \gamma $ channels. They are respectively given by
\be
\Gamma_{e^+e^-}  = {m_S \over 8 \pi} g_e^2\left(1-{4 m_e^2 \over m_S^2}\right)^{3/2}.
\ee
and 
\be
\Gamma_{\gamma \gamma}&=& {\alpha^2 m_S^3 \over 64 \pi^3}\left|\sum_{\ell=e,\mu,\tau}{g_\ell \over m_\ell}\tau_\ell\left[1+(1-\tau_\ell) f(\tau_\ell)\right]\right|^2, 
\ee
where $\tau_\ell \equiv {4 m_\ell^2/m_S^2}$ and the loop function $f(\tau)$ reads,  
\be
 f(\tau) = \left\{ \begin{array}{ll}
{\rm arcsin}^2\left( {\tau^{-1/2}} \right), & \hspace{0.5cm} \tau > 1 \\
-\frac{1}{4} \left[ \ln\left( \frac{1+\sqrt{1-\tau}}{1-\sqrt{1-\tau}} \right) -i\pi \right]^2, & \hspace{0.5cm} \tau\leq 1
\end{array}\right.\,.
\ee
For model A, the decay is dominantly through $S\to e^+ e^-$,
\beq
\Gamma_S^\text{A} = \Gamma_{e^+ e^-}+\Gamma_{\gamma\gamma} \approx \Gamma_{e^+ e^-}
\eeq 
For model B, the only decay channel is $S\to \gamma \gamma$ via $\mu$ loop, i.e.,
\beq
\Gamma_S^\text{B}= {\alpha^2 m_S^3 \over 64 \pi^3}{g_\mu^2 \over m_\mu^2}\left|\tau_\mu\left[1+(1-\tau_\mu) f(\tau_\mu)\right]\right|^2
\eeq

Given $\Gamma_S$, the decay length of the scalar $S$ is expressed by
\be
\label{decaylength}
L_{S}={E_S \over m_S}   \frac{\beta_S}{\Gamma_{S}},
\ee
 where $E_S$ is the energy of the scalar and the boost factor $\beta_S=\sqrt{1-{m_S^2/ E_S^2}}$. In particular, taking a fiducial choice of parameters, and the energy of the scalar $ E_S= 3$ GeV, we find
\begin{eqnarray} 
L_S =25~ {\rm cm} \times \left( \frac{5\times 10^{-4}}{g_\mu}\right)^2\times \left( \frac{100~{\rm MeV}}{m_S}\right)^2,~~~~{\rm Model~A},\\
L_S =20~ {\rm m}\times \left( \frac{5\times 10^{-4}}{g_\mu}\right)^2\times \left( \frac{100~{\rm MeV}}{m_S}\right)^4,~~~~{\rm Model~B}.
\end{eqnarray} 
These sizable decay lengths give a good motivation to search for $S$ in the beam-dump experiments.

A previous study of the muon-beam-initiated emission of axion-like particles, \cite{Essig:2010gu}, shares several common features 
with our scalar model. Light dark vector particles emitted from a muon beam, in the context of the NA64 experiment,  have been also studied in \cite{Gninenko:2014pea}. Our paper aims to extend 
these previous works to the scalar case, and explore the sensitivity reach on the $ m_S$ --$g_\mu$
parameter space.

\section{Signals and backgrounds}
\label{signalsandbackgrounds}
\subsection{Signals}
\subsubsection{NA64-type muon beam-dump experiment}
In this subsection we investigate the constraints from the NA64-type experiment at CERN.
NA64 is a fixed-target experiment searching for dark sector particles and kaon invisible decays at the CERN Super Proton Synchrotron (SPS).
The detailed setup of the NA64 experiment can be found in Refs. \cite{Andreas:2013lya,Gninenko:2014pea}, and the experiment has reported its first results from the $2.75\times 10^9$ electrons on target in 2017~\cite{Banerjee:2016tad}.
We adopt similar geometries of the target and detector as suggested in Refs. \cite{Andreas:2013lya,Gninenko:2014pea} and sketch the setup in Fig.~\ref{fig:setup}.  The target is made of lead (Pb) with a thickness of $\sim$ 20 cm. The length of the detector is $\sim$ 5 meters. As pointed out in Ref.~\cite{Gninenko:2014pea}, the muon beam has a maximum luminosity of $10^6$ muons per second in order to evade loss of the signal efficiency due to the pileup effect. We assume a three-month run of the experiment at the maximum luminosity,
which yields $8\times 10^{12}$ muons in total on target.  The incident muon beam energy, $E_{\mu, \text{beam}}$, is assumed to be around 160 GeV.

To estimate the reach, we need to calculation of the differential cross section of the beam-dump process  $\mu^+ + N \to \mu^+ + N + S$ as shown in Fig.~\ref{fig:brem}. 
Given the large muon beam energy of NA64, we use the improved Weizsacker-Williams (IWW) approximation~\cite{Kim:1973he} in the evaluation.

When  the beam energy is far greater than the mass of beam particle and the mass of produced particle, $E_\text{beam}\gg m_\text{beam}, m_S$, virtual photons generated by the highly-boosted beam particle are nearly transverse and behave as  plane waves. Then the virtual photon can be approximate as a real photon, and the phase space integration of a $2\to 3$ ($\mu^+ +  N \to \mu^+  + N + S$) process is simplified to a $2\to 2$ ($\mu^+  + \gamma \to \mu^+ + S$) process. This is the so-called Weizsacker-Williamsuses (WW) approximation~\cite{Kim:1973he}. The WW approximation can be further refined by using the fact that the production of $S$ is dominantly collinear when the energy of $S$ approaches the energy of $\mu$. This yields the improved  Weizsacker-Williamsuses (IWW) approximation where the integration limits on the virtuality, $t$, are further simplified~\cite{Kim:1973he, Bjorken:2009mm}. The original  IWW derivation~\cite{Kim:1973he, Bjorken:2009mm} regards the beam particle to be massless. A detailed derivation of IWW with massive beam particles is presented in~\cite{Liu:2016mqv}.  The resulting differential cross section is given by 

\be
\f{\d }{\d x} \sigma(\mu^+ + N \to \mu^+ + N + S) \simeq \f{g_\mu^2 \alpha^2}{12\pi} \chi \beta_\mu \beta_S \f{x^3\left[m_\mu^2 (3x^2-4 x+4)+2 m_S^2 (1-x)\right]}{\left[m_S^2(1-x)+m_\mu^2 x^2\right]^2},
\ee
where $x\equiv E_S/E_\mu$ is the ratio between the energy of the exotic scalar, $E_S$, and the energy of the muon, $E_\mu$, inside the material. The boost factor for muon and the new scalar are given respectively by $\beta_\mu = \sqrt{1-m_\mu^2/E_\mu^2}\approx 1$ and  $\beta_S = \sqrt{1-m_S^2/(x E_\mu)^2}$.  The effective photon flux, $\chi$, is given by
\be
\chi = \int_{t_{\min}}^{t_{\max}} \d t  \f{t-t_{\min}}{t^2} G_2(t)\simeq \int_{m_S^4/\left(4 E^2_\mu\right)}^{m_S^2+m_\mu^2} \d t  \f{t-m_S^4/\left(4 E^2_\mu\right)}{t^2} G_2(t),
\label{chiex}
\ee
where $G_2$ is the combined atomic and nuclear form factor. The explicit expression of $G_2$ is given in the appendix~\ref{appendix}.  The simplified integration on $t$ from the IWW approximation is implemented in the second approximation of \eqref{chiex}.

\begin{figure}[t]
\centering
\includegraphics[width=0.45\columnwidth]{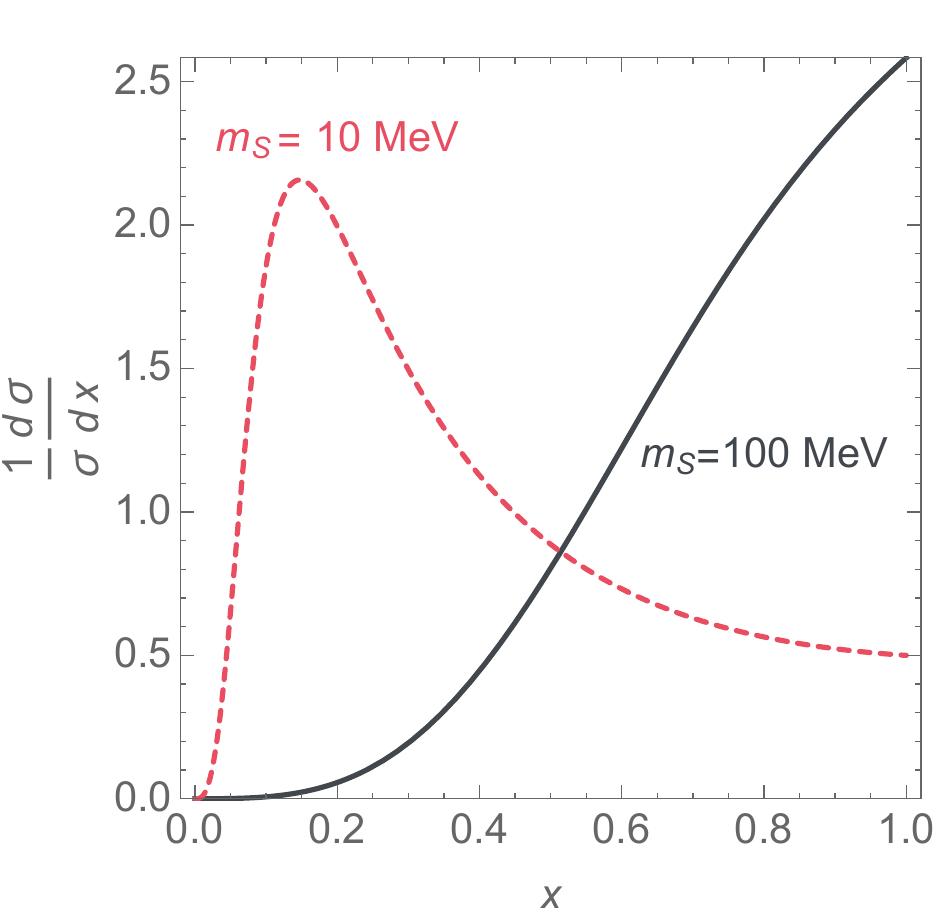}
\caption{\label{fig:x} Distributions of the fraction of the energy of the incident muon taken by the emission of an 
exotic scalar $S$ for various scalar masses. The red dashed curve is for $m_S=10$ MeV and the black solid  curve is for $m_S=100$ MeV. }
\end{figure}

Unlike earlier studies for the dark sector searches via electron beam-dumps, here we keep the beam particle mass to be non-zero.  As a consequence, 
the expected energy spectrum of $S$ varies quite significantly as $m_S$ increases. In 
Fig. \ref{fig:x}, we demonstrate this point by comparing $x$ spectrum for $m_S=10$ MeV and $m_S= 100$ MeV. 
The emission of light particles ($m_S \ll m_\mu$) clearly favors the low $x$ region, while for 
$m_S \approx m_\mu$, the outgoing $S$ takes more significant portion of the muon energy. 

The total number of events, $N_S$, can be obtained as a convolution of the 
production cross section and the decay probability. It is given by
\begin{align}
N_S =& N_{\mu} \int_{y_{\min}}^{y_{\max}} \d y\, n_{\rm atom} \int_{x_{\min}}^1 \d x {\d\sigma_{2\to 3}\over \d x}   \int_{z_{\min}}^{z_{\max}} \d z P(z)\\
=& N_{\mu} \int_{y_{\min}}^{y_{\max}} \d y\, n_{\rm atom} \int_{x_{\min}}^1 \d x {\d\sigma_{2\to 3}\over \d x}   \left(e^{-{z_{\min}\over L_S}}-e^{-{z_{\max}\over L_S}}\right),
\label{Ns}
\end{align}
where $N_\mu$ is the total number of incident muons and $\sigma_{2\to 3}$ is a shorthand for $\sigma(\mu^+ + N \to \mu^+ + N + S)$. $n_{\rm atom}$ is the number density of target nuclei and its integration over the muon penetration length, $y$, accounts for the number of target nuclei that a incident muon encounters. $P(z)$ stands for the decay probability density (per length) of $S$ decaying within the fiducial decay range from $z_{\min}$ to $z_{\max}$. $L_S$ is the decay length given by  Eq.~(\ref{decaylength}) with  $E_S = E_\mu x$. We also impose a $x_{\min}$ in the integration over $x$ to satisfy the specific requirements for the search of the signal.

Muon beams penetrate the target to a much longer depth compared to that of the electron beams.  It can loose energy through multiple mechanisms~\cite{Olive:2016xmw}. For the energy range we are interested in, from a few GeV to $\sim$100 GeV,  the muon energy loss is dominantly through the ionization and the stopping power $\langle \d E_\mu/\d y \rangle$  is relatively flat  with respect to the muon momentum~\cite{Olive:2016xmw}.  
Hence we approximate the muon energy loss per unit length $\langle \d E_\mu/\d y \rangle$ to be a \emph{constant}.  Consequently  $E_\mu$ is related to the penetration length $y$ via
\be
\Delta y \equiv y-y_{\min} ={E_{\mu, \text{beam}}-E_\mu \over \langle \d E_\mu/\d y \rangle},
\label{linear}
\ee
where $E_{\mu, \text{beam}}$ is the initial energy of the incident muon beam. \eqref{linear} can be used to simplify Eq.~(\ref{Ns}) into
\beq
N_S \approx  \frac{ N_\mu n_{\rm atom}}{\langle \d E_\mu/\d y \rangle} \int_{E_{\mu, \min}}^{E_{\mu, \text{beam}}} \d E_\mu  \int_{x_{\min}}^1 \d x  {\d\sigma_{2\to 3} \over \d x}  \left(e^{-{z_{\min}\over L_S}}-e^{-{z_{\max}\over L_S}}\right),
\label{simpleNs}
\eeq
where  the fiducial range for the decay, $z_{\min}$ and $z_{\max}$, are respectively given by 
\be
z_{\min} (E_\mu)= L_\text{tg}+L_\text{det} - \Delta y (E_\mu), \quad z_{\max} = \infty.
\ee
$L_\text{tg}$ and $L_\text{det}$ represent the lengths of the target and detector, respectively.  
For a lead target, $n_{\rm atom}=3.3\times 10^{22}$/cm$^3$ and $\langle \d E_\mu/\d y \rangle \approx 12.7\times 10^{-3}$GeV/cm for the relevant beam energy range~\cite{pdgmat}. For the thin target ($L_\text{tg}=0.2$ m), the muon energy after penetrating the target, $E_{\mu, \min}=159.75$ GeV, is close to the incident beam energy $E_{\mu, \text{beam}}=160$ GeV.  

The signature of the signal at NA64-type experiment for the incoming electron mode 
is a single EM shower in the target corresponding to the final state electron with significant missing energy. The required missing energy, $E_\text{miss}$, should be above expected background values. In the muon mode, the detection strategy would be modified somewhat, as one would need to detect the energy of the final state muon.  As suggested by~\cite{Gninenko:2014pea,Banerjee:2016tad}, here we require $E_{\rm miss} >E_{\mu,\rm bearm}/3\approx 53$ GeV, which is equivalent to setting $x_{\min}=1/3$ given $E_{\mu}$ is close to $E_{\mu, \text{beam}}$ for the thin target of NA64. We further assume the efficiency for the signal reconstruction $\approx 100\%$ and leave a more detailed detector modeling and study for the future.

\subsubsection{Fermilab muon beam-dump experiment}
Fermilab have capabilities of producing a more intense source of muons, albeit at a smaller muon beam energy. 
We suggest the simplest muon beam dump experiment, where a muon beam is fully stopped in a 
dense target with  a thickness of several meters.
Similar to the NA64-type of experiments, we can estimate the number of signal events using existing setup for the Fermilab 
 muon beam.
The incident muon beam energy we propose for the experiment, $E_{\mu, \text{beam}}$, is $\sim3$ GeV, as the accelerator complex is already tuned to 
this energy for the muon $g-2$ experiment~\cite{Chapelain:2017syu}. Such a beam will be completely stopped in 1.5 m thickness tungsten target ($n_{\rm atom}=6.3\times 10^{22}$/cm$^3$ and $\langle \d E_\mu/\d y \rangle \approx 22.1\times 10^{-3}$GeV/cm for the beam energy range~\cite{pdgmat}). Hence we propose a setup for the Fermilab muon beam-dump experiment as shown in Fig.~\ref{fig:setup}. The lengths of the target  (tungsten) and detector are around 1.5 m and 3 m, respectively. 
The total exposure taken for this proposed experiment is $10^7$ muons per second for 1 year of running, or 
$3\times 10^{14}$ muons in total on target. 

To avoid the background from soft muons, we adopt a lower limit on the muon beam energy $E_{\mu, \min}=0.5$ GeV.  
To estimate the number of signal events, we need to 
account for the muon energy loss inside the target via an integration over $E_\mu$ from 0.5 GeV to 3 GeV. Here the IWW approximation is not applicable since the muon energy around $E_{\mu,\min}$ is not much greater than $m_\mu$. Instead we use \texttt{MadGraph 5 aMC@NLO}~\cite{Alwall:2014hca} to obtain the cross sections of $\sigma(\mu^+ + N \to \mu^+ + N + S)$ for various $E_\mu$. The combined atomic and nuclear form factor, $G_2(t)$, is implemented in the model file (see Appendix~\ref{appendix} for more details) and the decay probability is implemented by reweighting the generated events. The procedure yields numerical values of the reweighed cross section 
\beq
\tilde \sigma (E_\mu) = \int_{x_{\min}}^1 \d x  {\d\sigma_{2\to 3} \over \d x}  \left(e^{-{z_{\min}\over L_S}}-e^{-{z_{\max}\over L_S}}\right)
\eeq
as a function of $E_\mu$. The fiducial range of the decay here is given by
\beq
z_{\min} (E_\mu)= L_\text{tg} - \Delta y (E_\mu), \quad z_{\max} (E_\mu)= L_\text{det} +z_{\min}.
\eeq
To estimate the number of signal events,  we interpolate over the samplings of $\tilde \sigma (E_\mu)$ and preform the integration over $E_\mu$ according to \eqref{simpleNs}.

The signature of the signal at Fermilab experiment is a decay with a sizable displaced vertex reconstructed from $e^+e^-$ or $\gamma \gamma$. A typical electron or photon tracker/calorimeter requires a minimum momentum/energy of the particle around 10 MeV.  This threshold is much smaller than the momentum of the decayed electrons or photons in the lab-frame, $p_{e,\gamma}^\text{lab}\approx \gamma_S m_S/2= x E_\mu/2$, and hence can be easily satisfied given a small $x$.  Therefore we approximate $x_\text{min}\approx 0$ in the estimation of $N_S$. Like the NA64 case, we further assume the efficiency for the signal reconstruction $\approx 100\%$ and leave a more careful detector modeling for future experimental studies.

\subsubsection{Difference between setups of NA64-type and Fermilab muon beam-dump experiments}

\begin{table}[htbp]
   \centering
   \begin{tabular}{@{} lcc @{}} 
      \hline
           & NA64-type & Fermilab\\
      \hline
      Incident muon beam energy, $E_{\mu, \text{beam}}$ &160 GeV & 3 GeV\\
      Total number of incident muons, $N_\mu$ & $8\times 10^{12}$ & $3\times 10^{14}$ \\
      Target material  &  Lead (Pb) & Tungsten (W)      \\
       Atomic number density, $n_\text{atom}$ & $3.3\times 10^{22}$/cm$^3$ & $6.3\times 10^{22}$/cm$^3$\\
 	Muon energy loss per unit length, $\langle \d E_\mu/\d y \rangle$ & $12.7\times 10^{-3}$~GeV/cm & $22.1\times 10^{-3}$~GeV/cm \\
	Target Length, $L_\text{tg}$  & 0.2 m  & 1.5 m \\
	 Detector Length, $L_\text{dec}$ & 5 m & 3 m\\
	 Min fiducial range for the decay, $z_{\min}$ & $L_\text{tg}+L_\text{dec}-\Delta y(E_\mu)$ &$L_\text{tg}-\Delta y(E_\mu)$ \\
	 Max fiducial range for the decay, $z_{\max}$ &  $\infty$ & $L_\text{tg}+L_\text{dec}-\Delta y(E_\mu)$ \\
      \hline
   \end{tabular}
   \caption{Parameters for the proposed muon beam-dump experiments at NA64 and Fermilab.}
   \label{twoexp}
\end{table}

\begin{figure}[t]
\centering
\includegraphics[width=0.45\columnwidth]{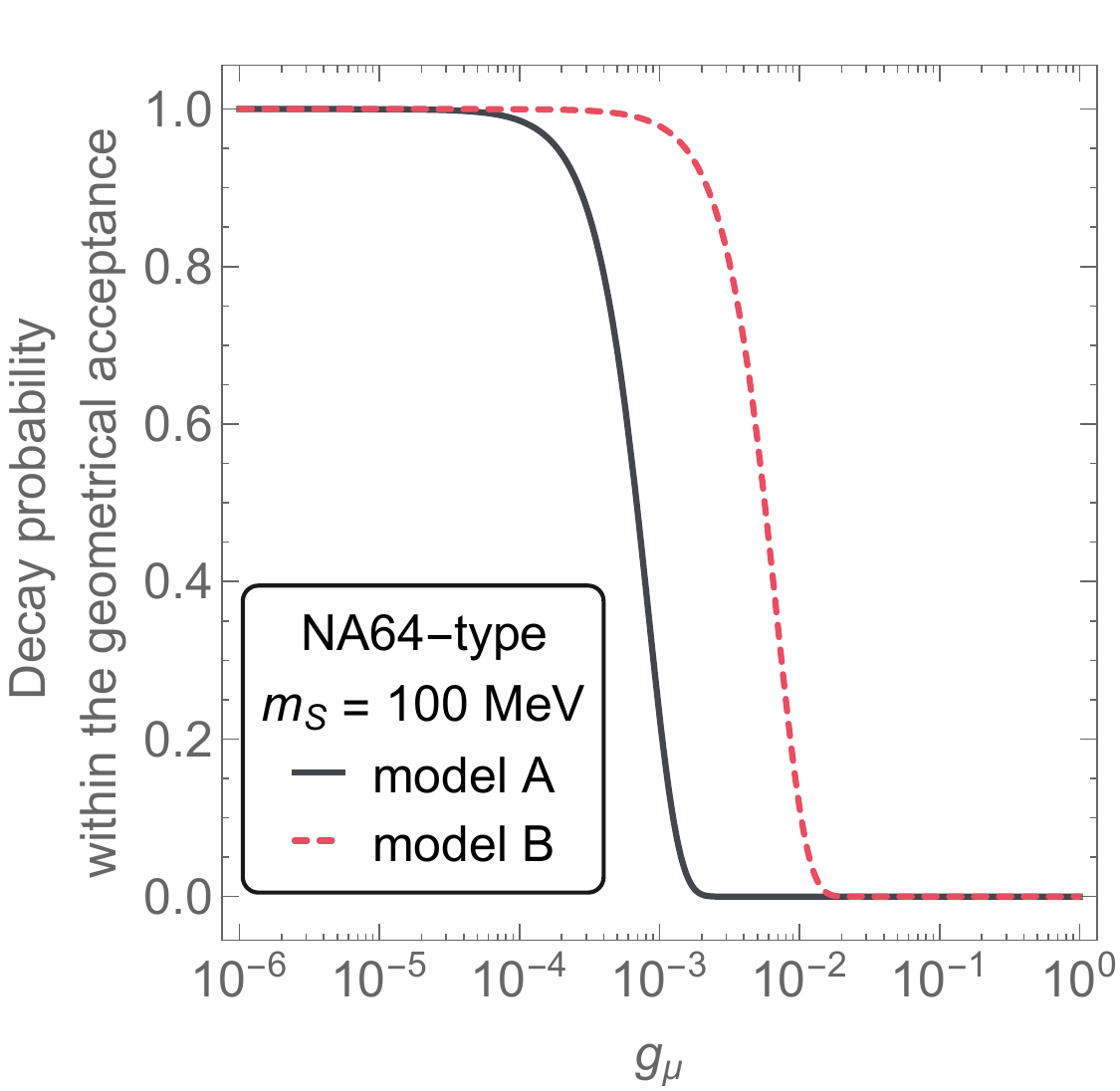}
\includegraphics[width=0.45\columnwidth]{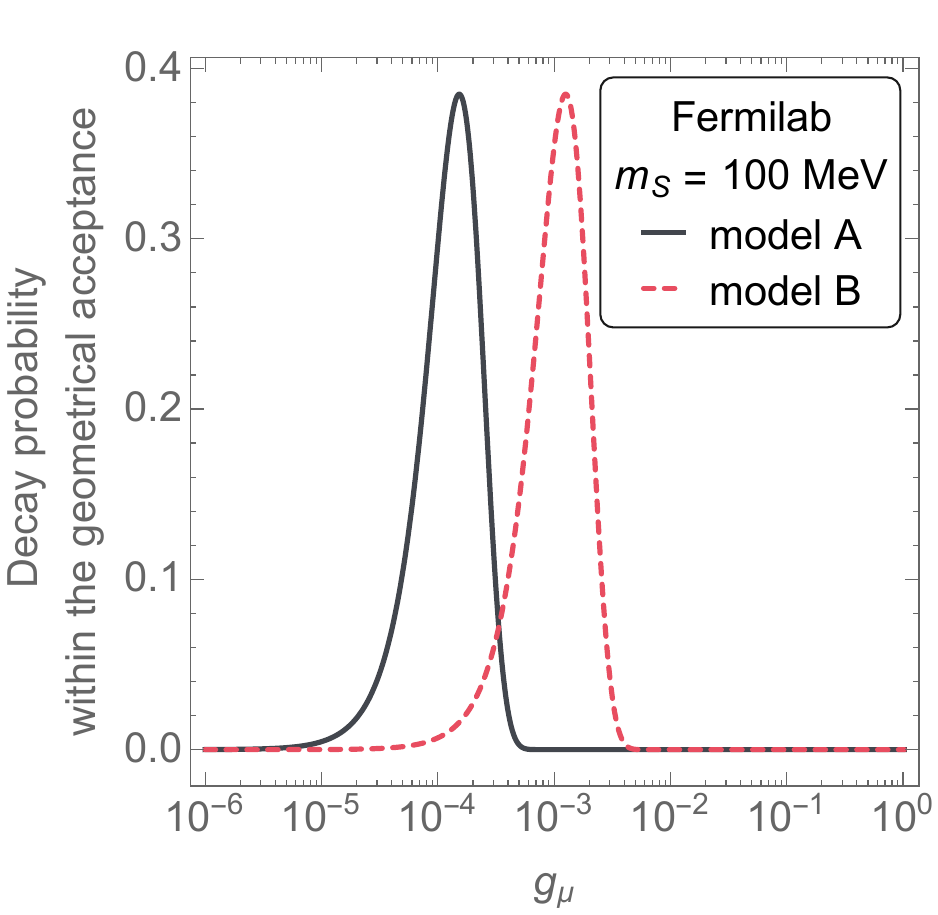}
\caption{\label{fig:decyprob} Decay probabilities of $S$ within the geometrical acceptance with $m_S=100$ MeV in model A (black) and model B (red dashed) at NA64 (left) and Fermilab (right), respectively.}
\end{figure}

Tab.~\ref{twoexp} summarizes the setups of NA64-type and Fermilab muon beam-dump experiments.
To illustrate the kinematic difference between the two setups, in
Fig.~\ref{fig:decyprob} we show the decay probabilities of $S$ within the geometrical acceptance with $m_s=100$ MeV in model A (black) and model B (red dashed) for the  NA64 (left) and Fermilab (right), respectively. For the NA64-type experiment (Fig.~\ref{fig:decyprob}, right panel), the curve corresponding to model B (red dashed) rises earlier as we gradually turn off the coupling. This is due to the fact that the decay length of $S$ in model B is much longer compared to that in model A.  Since the fiducial volume is from $\sim5$ m to infinity, we find that the decay probabilities remain close to one for very small $g_\mu$,  corresponding to the region of parameter space where the scalars are very long-lived. This tells us that the lower contours of NA64 in Fig.~\ref{fig:gmu_ms} is set by the production rate. 
Since the production rate of the light scalar is identical in both models A and B as illustrated in Fig.~\ref{fig:brem}, the lower limits for both models for NA 64 in Fig.~\ref{fig:decyprob} are the same.
For the muon beam-dump experiment at Fermilab, the width of the peak is due the finite size of the detector. Similar to the NA64 case, the curve corresponding to model B (red dashed) peaks earlier as the decrease of the coupling $g_\mu$. 

\subsection{Potential backgrounds}
The NA64 experiment has addressed the issues of potential background for the missing energy/momentum 
search, and many components of the experiment are tuned to reject various sources of backgrounds~\cite{Gninenko:2014pea, Banerjee:2016tad}. As described above, those studies suggest a missing energy cut $E_\text{miss}\gtrsim 50$ GeV, which is equivalent to requiring $1/3<x<1$. 
A more detailed analysis on potential backgrounds requires knowledge of the detector and is beyond the scope of this paper.

Next, we focus on possible backgrounds to the proposed Fermilab muon beam-dump experiment. 
All charged particles can be efficiently stopped inside the target given the target length we adopted. Potential backgrounds  may arise from the neutral long-lived kaons. They are produced by muons and  decay  after reaching the detector. They can mimic the long-lived scalars $S$. For  model B, 
$K_L \to \pi^+ \pi^- \pi^0$ and $K_L \to 3\pi^0$ decays are particularly 
worrisome, as they produce photon-like energy deposition.  
It is instructive to estimate how many $K_L$ reach the end of the target without suffering the degradation in 
energy. To do that we consider the muon-initiated kaon production cross section
$\sigma(\mu+N \to \mu + K^0 + X)$, where $X$ is 
a baryonic state with an open strangeness. The cross section size can be estimated using the WW approximation, and related to the underlying kaon photoproduction 
cross section,
\be
\d\sigma(\mu+N \to \mu + K^0 + X)  \sim \sigma(N +\gamma \to K^0 + X) \d n_\gamma,
\ee
where $n_\gamma$ is the number of quasi-real (or equivalent) photons carried by the muon.
The total number of $K_L$  is then given by
\be
\label{NsK}
N_\text{kaon} &\simeq &  \frac{N_\mu}{2}\int_{E_{\mu,\min}}^{E_{\mu, \text{beam}}} dE_\mu \frac{n_{\rm atom}}{\langle dE_\mu/dy \rangle} \int_{\omega_{\min}}^{\omega_{\max}} \sigma(\omega) dn_\gamma (\omega)\;e^{-{z_{\min}\over L_N}},
\ee
where $\sigma(\omega)$ is the kaon photoproduction cross section with an incident photon energy $\omega$ on a tungsten nucleus. We assume 
that the cross section is given by an incoherent sum of the production cross section 
on individual nucleons, $\propto A \; \sigma_\text{kaon}$ with $A$ being the number of nucleons. In turn,  $\sigma_\text{kaon}$ receives contributions from 
several subprocesses, $
\sigma(\gamma+n\to K^0 + \Lambda)$, $\sigma(\gamma+p\to K^0 + \Sigma^+) $, 
and $\sigma(\gamma+n\to K^0 + \Sigma^0).$
Each of these individual cross sections is about 0.8 $\mu$b on average, 
and becomes very small for $\omega < 1.5$~GeV \cite{Goers:1999sw,Tsukada:2007jy,Kaon}. 
The neutral kaons are produced predominantly in the upstream part of the target, and then 
propagate through about a meter of dense material. In the process of doing so their energy is degraded, 
and an idealized factor $e^{-{z_{\min}\over L_N}}$ in Eq. (\ref{NsK}) is to account for the probability of the produced kaons reaching the end of the target without interaction with the material. $L_N$ represents the nuclear collision length, which is $\simeq$ 6 cm for tungsten. 
Estimated that way, for one year of running, the number of kaons produced is around $\mathcal{O}(1)$ before any cuts. We expect that those background events can be further rejected by applying  selection 
criteria, such as absence of charged pions and/or invariant mass cuts. The kaon background can also be reduced by lowering the energy of the incident muon beam. Therefore we neglect the kaon background for Fermilab muon beam-dump in the projection below. More detailed background estimations can be achieved via specialized simulations for a concrete experimental design.

\section{Results}
\label{results}
\begin{figure}[t]
\centering
\includegraphics[width=0.45\columnwidth]{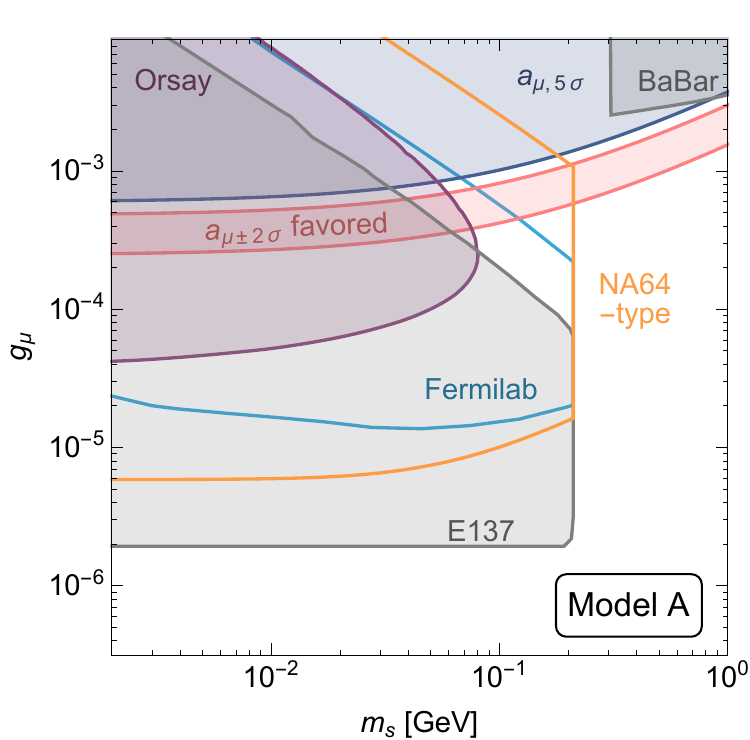}
\includegraphics[width=0.45\columnwidth]{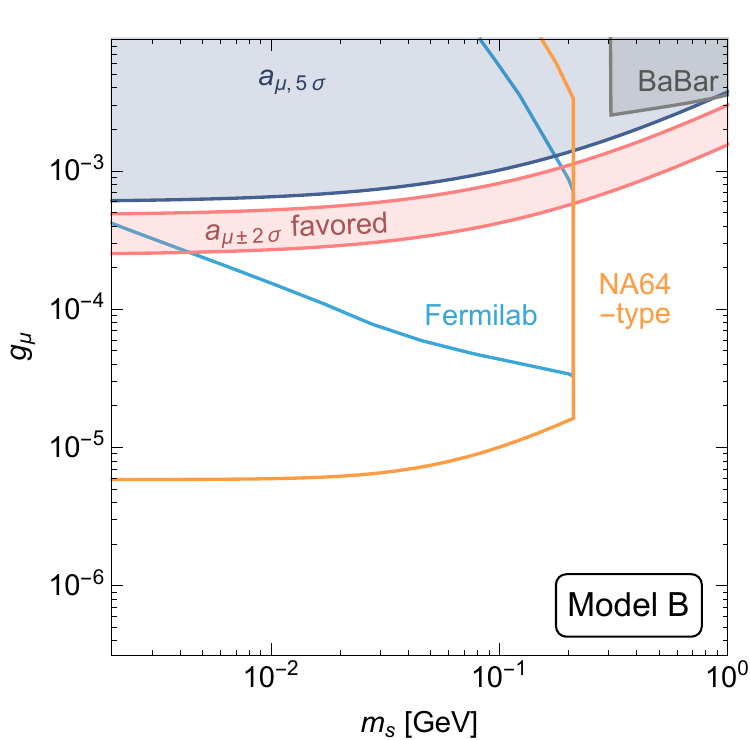}
\caption{\label{fig:gmu_ms} Prospects and constraints in the $m_S$ vs. $g_\mu$ plane for model A  (left) and model B  (right) respectively. The orange and cyan contours show the projected constraints from NA64-type and Fermilab muon beam-dump experiments respectively.
We include the  $2\sigma$ CL favored region and the $5\sigma$ CL  exclusions of $a_\mu$~\cite{Grange:2015fou,Mibe:2010zz}, and BaBar constraints~\cite{TheBABAR:2016rlg} for both models. For model A (left), we also include constraints from Orsay~\cite{Davier:1989wz} and E137~\cite{Bjorken:1988as}. See text for more details.
}
\end{figure}

Based on the estimations of  the number of signal events for the muon beam-dump experiments at NA64 and Fermilab,  we project sensitivities for the two proposed experiments for various models. 
Fig.~\ref{fig:gmu_ms} shows the resulting exclusions with 95\% confidence level (CL) on the $m_S$ vs. $g_\mu$ plane for model A  where $S$ dominantly decays through $e^+e^-$ (left) and model B where $S$ decays into $\gamma\gamma$ via a $\mu$-loop (right). The orange and cyan contours on the plots represent constraints from NA64 and Fermilab respectively. The projected constraints from the muon beam-dump experiments are compared with current constraints and the favored parameter space from other experiments. 
The red band is the favored region  to solve the $a_\mu$ discrepancy with 2$\sigma$ CL. The blue region are contributions of $S$ to $a_\mu$ that excluded with 5$\sigma$ CL. We have 
also show the limits from electron beam-dump experiments, Orsay \cite{Davier:1989wz} (purple) 
and E137 \cite{Bjorken:1988as} (gray) for model A. 
Those limits are not relevant for model B since there $S$ does not couple to electrons. The BaBar collaboration search through the process $e^+ e^- \to \mu^+ \mu^- S$  
\cite{TheBABAR:2016rlg} and excludes the upper right conner of the parameter space for both models. 

For model A, the projected constraints from NA64 and Fermilab is largely covered by existing E137 constraint for small $m_S$ and $g_\mu$.  Nevertheless, the muon beam-dump experiments will cover new grounds 
for the range of $m_S$ from 50 MeV to 210 MeV and $g_\mu$ from several of $10^{-5}$ to $10^{-3}$. They can effectively exclude the $a_\mu$ favored region for the $m_S$ range, which is not reached by E137. The importance of this region of parameter space has also been pointed out by \cite{Liu:2016qwd} in the
context of a more general model, where the authors attempt to address both the muon g-2 anomaly
 and the proton charge radius puzzle. As for model B, \emph{both} muon beam-dump experiments at Fermilab and NA64 can place strong limits and completely exclude the parameter space favored by muon $g-2$ experiment below the di-muon threshold. Again the NA64-type experiment is more sensitive to smaller $g_\mu$ region comparing to the Fermilab experiment. 
%

\section{Conclusions and Discussions}
\label{conclusions}

Muon beams have many applications in particle physics. Among fundamental physics applications, 
they have been used to study nuclear structure and perform precision measurements of $g-2$. 
The latter presents an intriguing 3$\sigma$--4$\sigma$ deficit of theoretical predictions relative to 
experimental observations. It could be a sign of low-mass new physics coupled to muons. All attempts to find such particles 
so far has rendered only exclusions on masses and couplings of such particles.  The majority of those searches have concentrated on hadron or electron-initiated production. In light of the main discrepancy coming from the muon sector, it makes 
sense to explore the possibility of light particles coupled predominantly to muons, and try to use a muon beam as 
a source of such particles. 

We have shown that muon beam-dump experiments at NA64 and Fermilab can effectively explore the light scalars that are predominantly coupled to muons.  Full UV-complete models with such scalars can be built. In this paper we have explored a simplified low-energy version of such models without going into details of the UV completion. The scalar $S$ can have a small, or vanishing, coupling to electrons. This also make the exotic scalar, $S$, long-lived, leading to the displaced decays in the beam-dump experiments. 
We have found that the experiments with muon beams indeed 
extend the reach to the parameter space of the exotic scalars. In particular, the favored parameter space to explain the $a_\mu$ discrepancy can be effectively probed.

Below we would like to discuss additional aspects of the low-mass new physics experiments with muon beams and the models they can explore:

\begin{itemize}

\item {\em Practical aspects of muon beam-dump at Fermilab.} The beam-dump experiment with the anomalous energy 
deposition downstream from the dump is among the simplest particle physics experiments. The muon beam energies
available at Fermilab allow to make this setup relatively compact, with the total length of a few meters. As such 
this proposed experiment could go into the $g-2$ experimental hall. Moreover, depending on the availability of protons, 
the proposed beam-dump can be run in parallel with the $g-2$ experiment. 

\item {\em Model dependence. } The simplified model of one scalar particle considered in this paper is an example of a 
physics goal that muon beam-dump experiments may pursue. An interesting variation of this is when the 
multiplicity of exotic states $N_d$ is large, as 
may occur in the models with extra dimensions where the dark forces 
are allowed to live \cite{McDonald:2010iq,Wallace:2013eya,Jaeckel:2014eba}, 
or in models with some conformal  dynamics, where the new states are continuously spread over the invariant mass \cite{Georgi:2007ek}. 
It is easy to see the qualitative difference in the phenomenology of such models compared to an exotic single state models. 
The effects of virtual dark force particles (such as corrections 
to $g-2$) can be enhanced by large multiplicity. 
Therefore, smaller individual couplings can be responsible for the same size of the corrections. Moreover, the 
mass step, $\Delta m_S$, can lead to overlapping resonances within a detector mass resolution, undermining the ``bump hunt" searches. 
This type of models with, {\em e.g.} a tower of dark photons,  will escape current direct searches at NA48/2, BaBar etc, but can be a source
of sizeable corrections in $g-2$. It is easy to see that such models generically lead to longer lifetimes of individual states, and therefore 
can be subjected to tighter displaced decay bounds. Such models can also be probed in the muon beam-dump experiments.

\item {\em The advantage of running NA64 in the muon mode. }  NA64 experiment currently occupies a 
unique niche (which can be followed up by a similar experiment in North America \cite{Izaguirre:2014bca}). In this paper we have argued
that a muon run in NA64 is warranted, as it provides a very strong sensitivity to models (model B) where the 
decay of $S$ happens well outside the detector. This adds to an important case of $L_\mu-L_\tau$ gauge boson 
with mass $m_{Z'} < 2 m_\mu$, where the final state of decay is always neutrinos  \cite{Gninenko:2014pea}. 

\item {\em Neutrino sources, SHiP.} 
In this paper, we have concentrated on considering dedicated experiments with muon beams.
Two other possibilities involve proton beam-dumps, which also creates a lot of muons, as well as 
beams of mesons used to source the neutrino beams. None of these possibilities is suitable for the 
missing energy or missing momentum studies. However, the anomalous energy deposition at the 
distance can indeed be probed, as is well known. Perhaps a very powerful probe of new physics coupled 
to muons can be achieved at a proposed SHiP facility \cite{Alekhin:2015byh}. There, a large number of muons created in the target 
propagates through tens of meters of material before getting stopped or deflected. The decay products of the 
light particles produced in the collision of muons with nuclei can be detected downstream, and a relatively 
short distance to detector (compared to past proton beam-dump experiments), as well as large boosts, 
may significantly increase the reach to unexplored parts of parameter space. Constraints on the muonic forces from the 
proton beam-dumps deserve a separate dedicated study.

\item {\em  Tau-initiated production. } Finally, model A with $g_\ell \propto m_\ell$ can be probed at 
high-luminosity $e^+e^-$ colliders through the process $e^+e^-\to \tau^+\tau^- S\to \tau^+\tau^- e^+e^-$ \cite{Batell:2016ove}. In light of the 
discussions in this paper, the decay of $S$ can also be displaced, producing a rather unique signature that is easy to be distinguished from the 
SM processes. Both Belle and BaBar collaborations could perform corresponding analyses.

\end{itemize}

\section*{Acknowledgements}

We would like to thank R. Essig,  E. Izaguirre, Y.-S. Liu, G. Magill, D. McKeen, R. Plestid, M. Schmaltz, and N. Toro for helpful discussions.  YZ thank N. Blinov, R. Essig, E. Izaguirre, V. Khachatryan, O. Mattelaer, A. Simonyan, and M. Takashi for setting up and testing the modified MadGraph 5 event generator for fixed-target experiments. The work of C.-Y.C and M.P. is supported by NSERC, Canada. 
Research at the Perimeter Institute is supported in part by the Government of Canada through NSERC and by the Province of Ontario through MEDT. Y.Z. is supported through DoE grant DE-SC0015845.

\appendix

\appendix

\section{Atomic and nuclear form factors}
\label{appendix}

The combined atomic and nuclear form factor, $G_2$, is presented in~\cite{Kim:1973he,Tsai:1973py,Tsai:1986tx, Bjorken:2009mm}. Two components contribute to $G_2$: (1) the elastic part is given by
\be
G_{2}^{\text{el}}(t) = \left(\f{a^2 t}{1+a^2 t}\right)^2 Z^2 \left(\f{1}{1+{t/d}}\right)^2 ,
\ee
where the virtuality $t$ represents the momentum transfer squared. $a=111 Z^{-1/3}/m_e$ under Thomas-Fermi model and $d=0.164~\text{GeV}^2A^{-2/3}$ where $A$ and $Z$ stand for the mass number and atomic number of the target material, respectively. $m_e$ is the electron mass~\cite{Kim:1973he,Tsai:1973py,Tsai:1986tx}; (2) the inelastic part,  in the limit $t/m_p^2$ is small, is given by 
\be
G_{2}^{\text{in}}(t)=\left(\f{a'^2 t}{1+a'^2 t}\right)^2 Z \left(\f{1+t(\mu_p^2-1)/(4m_p^2)}{\left(1+t/(0.71 \text{GeV}^2)\right)^4}\right)^2 ,
\ee
where $m_p$ is the proton mass, $a'=773 Z^{-2/3}/m_e$, and $\mu_p=2.79$~\cite{Kim:1973he}. 

For the parameter space of $m_{S}$ we are interested in, the expression for $G_{2}^{\text{in}}$ is valid, 
and we do not have to include the inelasticity at the nucleon level. 
The form factors dress the nucleus-nucleus-photon vertex with $G_2^{1/2}=(G_{2}^{\text{el}}+G_{2}^{\text{in}})^{1/2}$, which we implement in \texttt{MadGraph 5 aMC@NLO} according to~\cite{MG5FF}.

\newpage
\bibliography{ref}

\end{document}